\documentclass{jpsj-suppl}

\title{Angular Dependence of the High-Magnetic-Field Phase Diagram of URu$_2$Si$_2$}

\author{\name{Gernot W. \surname{Scheerer}}$^1$, \name{William \surname{Knafo}}$^1$, \name{Dai \surname{Aoki}}$^2$ and \name{Jacques \surname{Flouquet}}$^2$}

\inst{$^{1}$Laboratoire National des Champs Magn\'{e}tiques Intenses, CNRS-UPS-UJF-INSA, 143 avenue de Rangueil, 31400 Toulouse, France\\
$^2$Institut Nanosciences et Cryognie, SPSMS, CEA-Grenoble, 17 rue des Martyrs, 38054 Grenoble, France}

\abst{We present measurements of the magnetoresistivity $\rho_{xx}$ of URu$_2$Si$_2$ single crystals in high magnetic fields up to 60 T and at temperatures from 1.4 K to 40 K. Different orientations of the magnetic field have been investigated permitting to follow the dependence on $\theta$ of all magnetic phase transitions and crossovers, where $\theta$ is the angle between the magnetic field and the easy-axis \textbf{c}. We find out that all magnetic transitions and crossovers follow a simple 1/cos$\theta$-law, indicating  that they are controlled by the projection of the field on the \textbf{c}-axis.}

\kword{URu$_2$Si$_2$, magnetoresistivity, angular dependence, field-temperature-phase diagram}

\begin{document}

\maketitle

\section{Introduction}

URu$_2$Si$_2$ is one of the most recently studied heavy-fermion compounds\cite{mydosh}. It exhibits the so-called "hidden-order" state below $T_0= 17.5$~K \cite{palstra85,schablitz,maple,bourdarot,amitsuka}, for which the order parameter is still not identified, and has a transition to superconductivity at 1.5~K \cite{schablitz,maple}. The resistivity \cite{palstra86} and magnetic susceptibility \cite{palstra85,dawson} of URu$_2$Si$_2$ are strongly anisotropic, i.e., angle-dependent. The upper critical field $H_{c2}$ related to the destruction of the superconductivity is also strongly angle-dependent \cite{ohkuni99}. Under pressure a quantum phase transition leads to an antiferromagnetic ground state \cite{motoyama,amitsuka,hassingerPT}. A magnetic field $\mathbf{H}$ parallel to the easy-axis \textbf{c} induces a cascade of low-temperature magnetic phase transitions between 35~T and 39~T from the "hidden-order" state to a polarized paramagnetic sate. In the last 10 years, the $H$-$T$-phase diagram for $\mathbf{H} \parallel \mathbf{c}$ has been investigated by magnetization \cite{sugiyama99}, ultrasound \cite{suslov}, and resistivity \cite{kim03} measurements. The angle-dependent behavior of the low-temperature quantum phase transitions $H_1$, $H_2$ and $H_3$ (see Figure 1) at 1.5 K and up to 45 T is also known from magnetization \cite{sugiyama90} and magnetoresistivity \cite{jo} measurements: $H_1$, $H_2$ and $H_3$ follow a simple 1/cos$\theta$-law, where $\theta$ is the angle between the magnetic field and the $\mathbf{c}$-axis. Other angle-dependent measurements focused on the study of quantum oscillations \cite{ohkuni97,ohkuni99,keller,hassinger,shishido} at lower fields, i.e. inside the hidden-order phase, in order to investigate the Fermi surface. We present here a study of the angle-dependence of the magnetoresistivity of URu$_2$Si$_2$ in a larger experimental window, in high magnetic fields up to 60~T and temperatures up to 40~K. The complete \textit{H}-\textit{T}-phase diagram is investigated, including all transition and crossover lines.

\section{Experimental details}

The high-quality single crystals of URu$_2$Si$_2$ studied here have been grown by the Czochralski technique. All experiments were done in a $^4$He-cryostat and at ambient pressure. We have measured the electrical resistivity $\rho_{xx}$ ($\mathbf{I},\mathbf{U}\parallel\mathbf{a}$, where $I$ and $U$ are the current and voltage, respectively) as a function of the magnetic field using the four-contact method. The high quality of our URu$_2$Si$_2$ single crystals is shown by their residual resistivity ratio RRR~$=\rho_{x,x}(\rm{300K})/\rho_{x,x}(\rm{2K})$, where $\rho_{x,x}$ is the zero-field resistivity. The RRR are 90 for sample \#1 and 225 for sample \#2. Measurements have been done in 6mm-bore and 17mm-bore 60-T magnets of the LNCMI-T facility with a duration of the pulse of 150 ms and 300 ms, respectively. Sample \#1 has been measured in the configurations [$\mathbf{H}\parallel\mathbf{c}$; $\mathbf{I},\mathbf{U}\parallel\mathbf{a}$] and [$(\mathbf{H},\mathbf{c})=20^{\circ}$; $\mathbf{I},\mathbf{U}\parallel\mathbf{a}$]. Sample \#2 has been measured in a rotational probe permitting \textbf{H} to turn in the \textbf{a}-\textbf{c}-plane from transversal [$\mathbf{H}\perp\mathbf{I},\mathbf{a}$; $\theta=0^{\circ}$] to longitudinal [$\mathbf{H}\parallel\mathbf{I},\mathbf{a}$; $\theta= (\mathbf{H},\mathbf{c})=90^{\circ}$] configurations.

\section{Results}

\begin{figure}[tbh]
\begin{center}
\includegraphics[width=.95\linewidth]{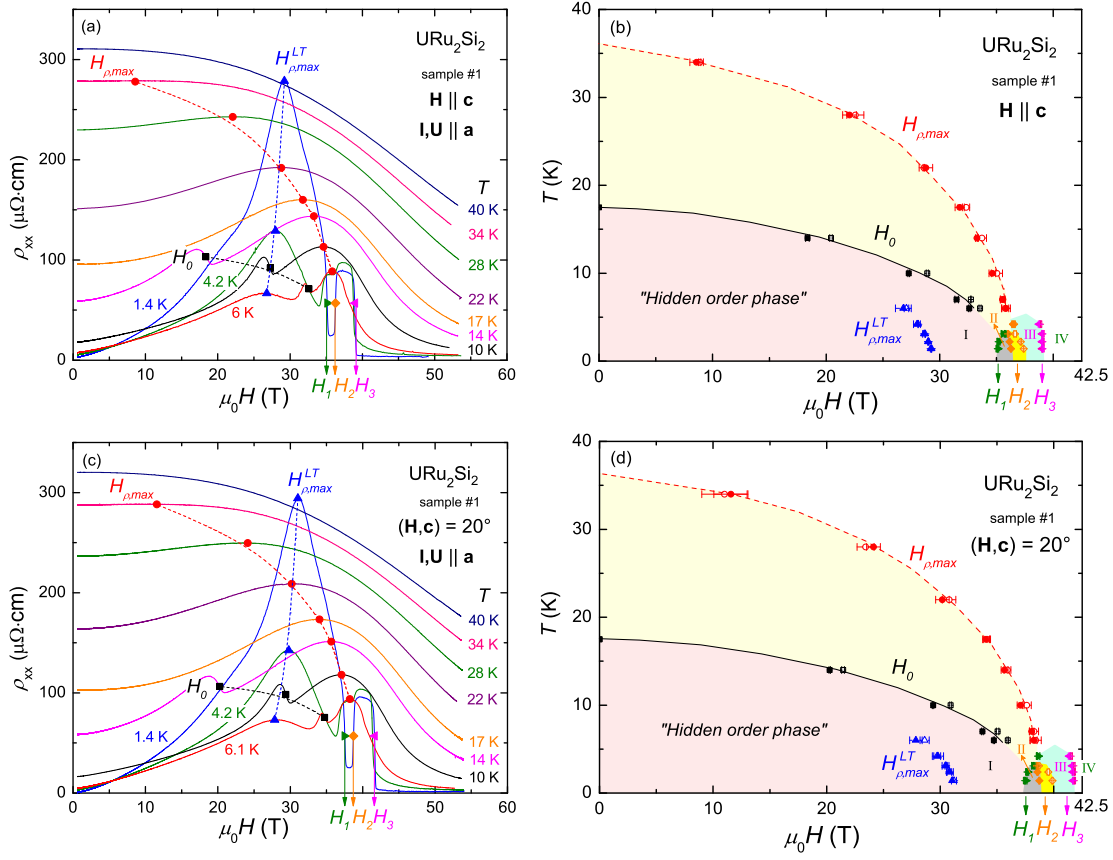}
\end{center}
\caption{ Left: Resistivity $\rho_{xx}$ as a function of the magnetic field $H$ of sample \#1 at temperatures from 1.4~K to 40~K, with (a)  $\mathbf{H} \parallel \mathbf{c}$ and (c) $(\mathbf{H},\mathbf{c})=20^{\circ}$. Symbols indicate the anomalies in the resistivity at the phase transition fields $H_0$ (black squares), $H_1$ (green triangles), $H_2$ (orange diamonds), and $H_3$ (magenta triangles), and crossover fields $H_{\rho,max}$ (red circles) and $H^{LT}_{\rho,max}$ (blue triangles). Dashed lines as guides to the eyes. Right: Resulting $H-T$-phase diagrams for (b) $\mathbf{H} \parallel\mathbf{c}$ and  (d) $(\mathbf{H},\mathbf{c}) = 20^{\circ}$. Full symbols correspond to increasing field and open symbols to decreasing field. The lines are guides to the eyes.}
\label{transLAYOUT}
\end{figure}

Figure \ref{transLAYOUT} (a) presents the resistivity $\rho_{xx}$ of sample \#1 as a function of the magnetic field \textbf{H} $\parallel$ \textbf{c} for different temperatures $T$ from 1.4~K to 40~K. Above 5~K the magnetoresistivity shows a phase transition at the field $H_0$, which is related to the destruction of the "hidden-order" phase, and a broad maximum at the field $H_{\rho,max}$, which is related to a high-temperature crossover, presumably controlled by the destabilization of intersite electronic correlations \cite{scheerer}. $H_0$ is defined at the extremum of $\partial\rho_{x,x}/\partial H$ and vanishes at 17~K, reaching 34~T at low temperature. The crossover field $H_{\rho,max}$ vanishes at 40~K and reaches 36~T at low temperature. Below 5~K the resistivity increases strongly with the magnetic field from 0 to 30~T. After a maximum at $H^{LT}_{\rho,max}$ ($\approx$ 30~T) the resistivity decreases until 35~T. Above 35~T the "hidden-order" phase is destroyed and a cascade of first-order quantum phase transitions is induced at the fields $H_1$, $H_2$ and $H_3$, which are leading to a polarized paramagnetic state above 39~T. At 1.5~K, $\mu_0H_1=35.1\pm0.1$~K, $\mu_0H_2=37.4\pm0.1$~T for increasing field and $36.3\pm0.1$~T for decreasing field, and $\mu_0H_3=39.0\pm0.1$~T are defined at the extrema of $\partial\rho_{x,x}/\partial H$. Figure \ref{transLAYOUT} (b) is the resulting $H$-$T$-phase diagram of URu$_2$Si$_2$ obtained from our magnetoresistivity data for \textbf{H} $\parallel$ \textbf{c}.

\begin{figure}[tbh]
\begin{center}
\includegraphics[width=.5\linewidth]{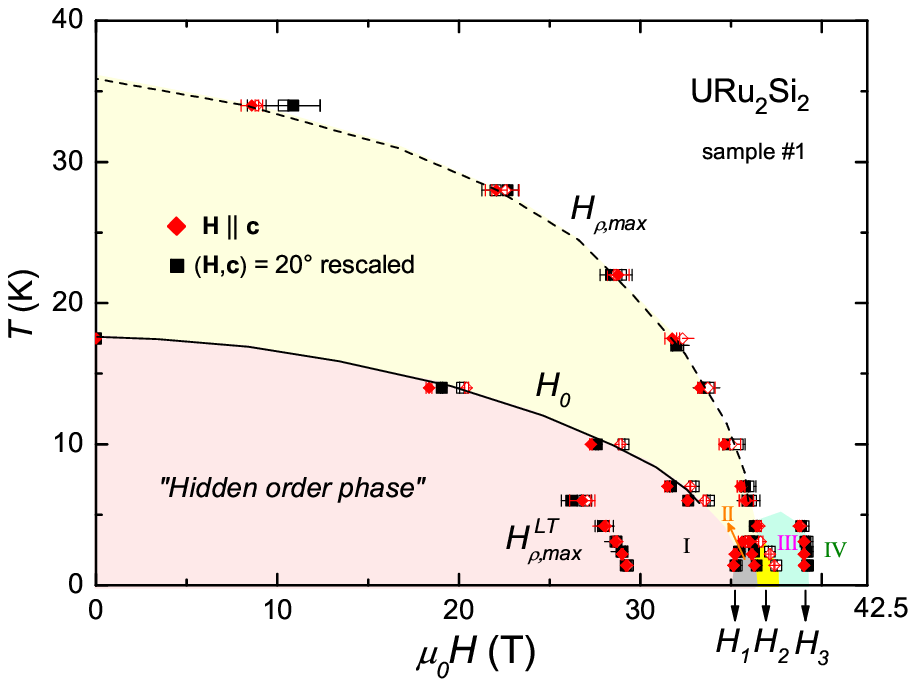}
\end{center}
\caption{$H$-$T$-phase diagram obtained for $(\mathbf{H},\mathbf{c})=\theta=20^{\circ}$ (black squares), superposed with a scaling factor $0.94 = \rm{cos}\theta$ with the phase diagram obtained for $\mathbf{H}\parallel\mathbf{c}$ (red diamonds).}
\label{transPHASEDIARESC}
\end{figure}

Figure \ref{transLAYOUT} (c) shows the resistivity $\rho_{xx}$ of sample \#1 as a function of the magnetic field when $\theta=(\mathbf{H},\mathbf{c})\approx20^{\circ}$. The general form of the magnetoresistivity is similar to that obtained for $\mathbf{H}\parallel\mathbf{c}$  (Figure \ref{transLAYOUT} (a)) showing the same transitions. The main difference is a shift of the resistivity curves to higher field values. For example the transition field $H_1$, which is related to the destruction of the "hidden-order" phase at 1.5 K, is shifted from 35.1 T to 37.6 T. Figure \ref{transLAYOUT} (d) is the resulting $H$-$T$-phase diagram of URu$_2$Si$_2$ obtained from our magnetoresistivity data for (\textbf{H},\textbf{c}) = 20$^{\circ}$. Figure \ref{transPHASEDIARESC} shows that the phase diagram obtained for $(\mathbf{H}\parallel\mathbf{c})=20^{\circ}$ is re-scalable on the phase diagram obtained for $\mathbf{H}\parallel\mathbf{c}$, with a scaling factor of 0.94 which corresponds approximately to cos$\theta$ with $\theta=20^{\circ}$. This indicates that the physics of the whole $H$-$T$-phase diagram  is mainly governed by the projection of the magnetic field along the \textbf{c}-axis.

\begin{figure}[tbh]
\begin{center}
\includegraphics[width=.95\linewidth]{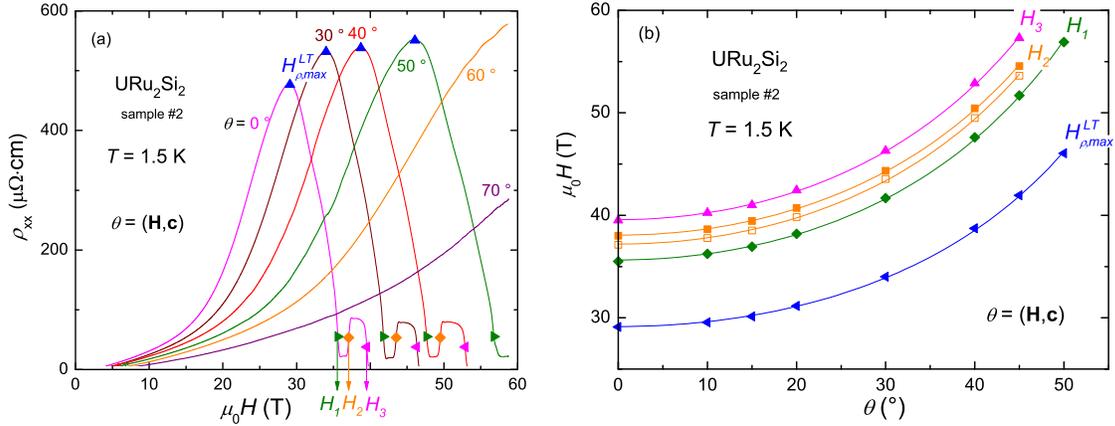}
\end{center}
\caption{ a) Resistivity $\rho_{xx}$ as a function of the magnetic field $H$ of sample \#2 at 1.5~K for different angles $\theta$ between the magnetic field \textbf{H} and the \textbf{c}-axis. Symbols indicate the transition fields $H_1$ (green triangles), $H_2$ (orange diamonds), $H_3$ (magenta triangles) and the crossover field $H^{LT}_{\rho,max}$ (blue triangles). By increasing $\theta$ the resistivity curves shift to higher field values with no change of their general form. For $\theta>50^{\circ}$ the anomalies are shifted out of the field range [0-60~T]. b) Resulting $H$-$\theta$-phase diagram made of the $\theta$-dependence of $H_1$, $H_2$, $H_3$ and $H^{LT}_{\rho,max}$. The solid lines represent 1/$\rm{cos}\theta$-fits to the data.}
\label{rotLAYOUT}
\end{figure}

Figure \ref{rotLAYOUT} (a) shows the magnetoresistivity $\rho_{xx}$ of sample \#2 at $T = 1.5$~K as a function of the magnetic field, for different angles $\theta$ between the magnetic field $\mathbf{H}$ and the \textbf{c}-axis. As emphasized in \cite{scheerer}, a strong sample-dependence of the magnetoresistivity at low temperature and below 35~T indicates that the signal is then controlled by an orbital effect. Here, the magnetoresistivity peaks at 30~T at $\approx500\mu\Omega.$cm in sample \#2 and at $\approx300\mu\Omega.$cm in sample \#1, this orbital signal being enhanced when the electronic mean-free path is higher, i.e., in sample \#2 where the RRR is higher. When $\theta$ increases, the magnetoresistivity shifts to higher field values without significant change in the general form of $\rho_{xx}(H)$. The maximum of $\rho_{xx}$ at $H^{LT}_{\rho,max}$ is slightly increasing, but the height of the plateau between $H_2$ and $H_3$ remains constant. The $\theta$-dependance of the transition fields $H_1$, $H_2$, $H_3$ and the crossover field $H^{LT}_{\rho,max}$ is represented in Figure \ref{rotLAYOUT} (b). They all follow a $1/\rm{cos}\theta$-law. This confirms that for $\theta$ up to 50~$^{\circ}$ the physics of those transitions and crossover only depends on the projection of the magnetic field on the \textbf{c}-axis.

\section{Discussion}

Our high-field magnetoresistivity measurements permitted to extend the study of the angle-dependence of the $H$-$T$-phase diagram of URu$_2$Si$_2$ to higher fields up to 60~T and higher temperatures up to 40~K (previous works \cite{sugiyama90,jo} have been performed below 1.5~K and up to 45~T). We confirmed that, at low temperature and up to 60~T, all magnetic transition fields $H_1$, $H_2$ and $H_3$ follow a $1/\rm{cos}\theta$-law, i.e. for angles $\theta$ between $\mathbf{H}$ and $\mathbf{c}$ varying up to 50~$^{\circ}$. As well, we have shown here that, up to 17.5~K, the magnetic transition line $H_0$ related to the destruction of the hidden-order also follows a $1/\rm{cos}\theta$-law, as well as the magnetic crossover $H_{\rho,max}$ which persists up to 40 K and which is associated to the development of intersite electronic correlations \cite{scheerer}. In addition, we have shown that the crossover field $H^{LT}_{\rho,max}$ at which the low-temperature magnetoresistivity is maximal also follows a $1/\rm{cos}\theta$-law. In Ref. \cite{scheerer}, this maximum has been related to a field-induced modification of the Fermi surface. The transition and crossover fields $H_0$, $H_1$, $H_2$, $H_3$, and $H_{\rho,max}$ related to the $f$-electron magnetic properties as well as the crossover field $H^{LT}_{\rho,max}$ related to a Fermi surface reconstruction are thus all controlled by the projection of the magnetic field along the $c$-axis, which is the easy magnetic axis of the system. This confirms that the magnetic properties of the $5f$-electrons and that of the Fermi surface are intimately connected in URu$_2$Si$_2$.

\section*{Acknowledgements}

We acknowledge J. B\'{e}ard, P. Delescluse, M. Nardone, C. Proust, and A. Zitouni for technical support and K. Behnia, F. Bourdarot, F. Hardy, H. Harima,H. Kusunose, J. Levallois, and C. Proust for useful discussions. This work was supported by the French ANR DELICE and by Euromagnet II via the EU under Contract No. RII3-CT-2004-506239.

\end{document}